\documentclass[preprint,12pt,review]{elsarticle}

\usepackage{graphicx}
\usepackage{epstopdf}
\usepackage{float}
\usepackage{subfig}
\usepackage{caption}
\usepackage{booktabs}
\usepackage{multirow}
\usepackage{makecell}
\usepackage{xurl}
\usepackage{amssymb}
\usepackage{amsmath}
\usepackage{array}
\usepackage{threeparttable}
\usepackage{bm}
\usepackage{hyperref}

\journal{Elsevier}

\begin{document}

\begin{frontmatter}

\title{Real-time Bidding Strategy in Display Advertising: An Empirical Analysis}

\author[uestc]{Mengjuan Liu\corref{cor1}}
\ead{mjliu@uestc.edu.cn}
\cortext[cor1]{Corresponding to: University of Electronic Science and Technology of China, No.4, Section 2, North Jianshe Road, Chengdu, 610054, China.}

\author[uestc]{Zhengning Hu}
\ead{202021090114@std.uestc.edu.cn}

\author[uestc]{Zhi Lai}
\ead{laizhi727@126.com}

\author[uestc]{Daiwei Zheng}
\ead{2018040701004@std.uestc.edu.cn}

\author[uestc]{Xuyun Nie}
\ead{xynie@uestc.edu.cn}

\address[uestc]{Network and Data Security Key Laboratory of Sichuan Province, University of Electronic Science and Technology of China, Chengdu, 610054, China}

\begin{abstract}
Bidding strategies that help advertisers determine bidding prices are receiving increasing attention as more and more ad impressions are sold through real-time bidding systems. This paper first describes the problem and challenges of optimizing bidding strategies for individual advertisers in real-time bidding display advertising. Then, several representative bidding strategies are introduced, especially the research advances and challenges of reinforcement learning-based bidding strategies. Further, we quantitatively evaluate the performance of several representative bidding strategies on the iPinYou dataset. Specifically, we examine the effects of state, action, and reward function on the performance of reinforcement learning-based bidding strategies. Finally, we summarize the general steps for optimizing bidding strategies using reinforcement learning algorithms and present our suggestions.
\end{abstract}


\end{frontmatter}


\section{Introduction}
\label{Introduction}
With the great success of the mobile Internet, online advertising has become the most crucial channel for brand promotion and merchandising \citep{choi2020online}. In 2021, Google’s ad revenue amounts to \$209.49 billion \citep{statista}. As one of the most striking advances in online advertising, real-time bidding (RTB) \citep{wang2015real} has received increasing attention since it improves the efficiency and transparency of the ad ecosystem \citep{yuan2013real}. In RTB, the publishing media sells ad impressions through public auctions, and advertisers bid on their targeting ad impressions in real-time and pay for their winning impressions. Therefore, it requires the bidding agent to make accurate user feedback predictions for each ad impression and determine a reasonable bidding price to maximize the long-term revenue \citep{grigas2017profit} of the ad campaign. \par
\begin{figure}[!]
	\centering
		\includegraphics[width=\textwidth]{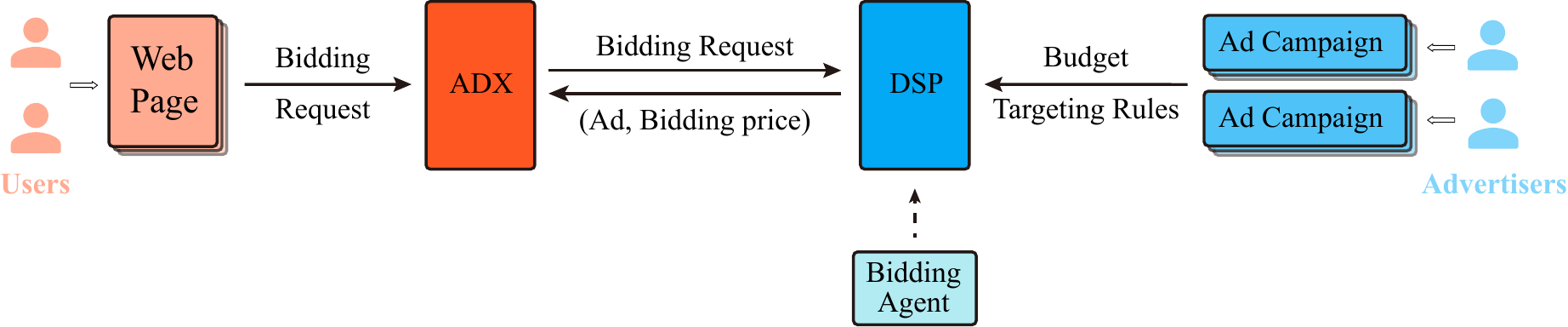}
	  \caption{An illustration of the bidding process in RTB display advertising}\label{fig1}
\end{figure}
Figure \ref{fig1} illustrates the entire process of an advertiser participating in bidding for an ad impression. Initially, the advertiser registers an ad campaign on the Demand Side Platform (DSP) and specifies the campaign’s budget as well as targeting rules for each ad delivery period (usually a day). Bidding agents running on DSP participate in RTB on behalf of advertisers. When a user browses a web page containing an ad slot, the script embedded in the ad slot initiates a bidding request to the ad exchange (ADX), which sends the bidding request to its interconnected DSPs. Then, bidding agents on the DSPs bid the impressions that meet ad campaigns’ targeting rules. The DSP feeds the highest internal bidding price to ADX, which determines the winner by the generalized second pricing (GSP) auction mechanism \citep{yuan2014survey}. Finally, the winning advertiser’s ad will be delivered to the user. Usually, DSP will track the user’s click or conversion behavior to optimize the advertiser’s bidding strategy and improve budget efficiency \citep{zhang2015statistical}. In RTB, the bidding agent typically determines the bidding price for each ad impression based on the campaign’s available budget, the auction environment, and the impression valuation. Therefore, the bidding strategy adopted by the agent plays an important role, as it dictates the revenue that the advertiser receives for a given budget. \par
Currently, bidding strategies fall into two main categories: static and dynamic. Static bidding strategies take the form of linear \citep{LIN} or non-linear \citep{ORTB} functions whose parameters are learned by heuristic algorithms from the data of historical ad delivery periods or optimization methods and are used in new ad delivery periods. Static bidding strategies are the most widely adopted on DSPs due to their simplicity and ease of deployment. Unfortunately, the RTB environment is highly dynamic, making it difficult for static bidding strategies to get the desired results in new ad delivery periods. A promising solution is to optimize the bidding strategy using reinforcement learning (RL) \citep{RL}, which shows excellent sequential decision-making capability. RL-based bidding strategies adjust the bidding function according to the RTB auction environment, thus ensuring ideal marketing results in the new ad delivery periods. Although RL-based bidding strategies have made some research progress, such as RLB \citep{RLB}, DRLB \citep{DRLB}, and FAB \citep{FAB}, they are still very far from commercial deployment. Because the state, action, and reward function designs of RTB auction environments are challenging, hindering the application of RL in optimizing RTB strategies. \par
This paper studies the bidding strategy for individual advertisers in RTB. We first give a formal description of the bidding strategy optimization problem and, in particular, introduce the objective of the problem in the reinforcement learning context. Then, we sort out representative bidding strategies, especially the difficulties in optimizing real-time bidding strategies using RL and recent research advances. On this basis, we quantitatively evaluate the performance of representative bidding strategies based on the benchmark dataset iPinYou \citep{IPINYOU}. Specifically, we discuss the impact of the design of the Markov Decision Process (MDP) components (state, action, and reward function) on the bidding performance in RL-based bidding strategies. Finally, we summarize the general steps for optimizing bidding strategies using RL algorithms and present our suggestions. \par
Although the intelligent bidding system is becoming a hot research topic in computing advertising because of its vast business value, the existing bidding systems still have much potential for improvement in adapting to the highly dynamic auction environment. The main contributions of this paper are summarized as follows:
\begin{itemize}
\item We provide a thorough investigation of representative bidding strategies and give an objective discussion of their flaws and advantages through quantitative analysis.
\item We propose a general framework for bidding systems to optimize their strategies in real-time by reinforcement learning, thus providing practical guidelines for developing intelligent bidding systems.
\item Finally, we have open-sourced all experimental codes on GitHub\footnote{\url{https://github.com/hzn666/RLBid\_EA}}, expecting that it will facilitate researchers to reproduce our bidding experiments quickly. To the best of our knowledge, this is the first comprehensive study of real-time bidding strategies from an empirical analysis perspective.
\end{itemize}
\par
The rest of this paper is organized as follows. Section \ref{Problem Definition} describes the optimization problem of real-time bidding strategies and the basic concepts of RL-based bidding strategies. In Section \ref{Representative Bidding Strategies}, we present representative bidding strategies and discuss the current challenges. Section \ref{Experiment Setting} gives the experimental setup of this paper. In Section \ref{Empirical Analysis}, we compare the performance of several typical bidding strategies and discuss the impacts of the design of MDP components on the performance of RL-based bidding strategies. Finally, Section \ref{Conclusion} concludes this paper and gives the general steps for optimizing the bidding strategy based on RL algorithms. \par
\section{Problem Definition}
\label{Problem Definition}
\subsection{Real-time Bidding Strategy}
\label{Real-time Bidding Strategy}
In RTB, DSP bidding agents represent advertisers bidding for ad impressions based on the bidding strategy. Whenever an ad impression that meets the advertiser’s targeting rules arrives at the DSP, the bidding agent calculates the bidding price based on the available budget and the valuation of the impression, where the valuation is typically the predicted click-through rate (pCTR) \citep{chapelle2014simple} for the impression. If the advertiser wins the auction, it needs to pay for this winning impression. GSP auction mechanism is now commonly used in RTB systems, which means that the winner only pays the second-highest price (market price) in this auction. Once the budget runs out, the bidding agent will stop bidding. Therefore, the goal of optimizing bidding strategy for individual advertisers is to maximize revenue within one ad delivery period for a given budget.
\begin{equation}
\begin{split}
bid^*(i) &=\underset{bid(\cdot)}{\mathop{\arg \max}}\sum\limits_{i \in I} w(bid(i),\ i) \times v(i) \\
s.t. &\sum\limits_{i\in I} w(bid(i),\ i) \times cost(i) \le B
\end{split}
\label{equ1}
\end{equation}
\par
We formulate the bidding strategy optimization problem as an optimization problem with constraints as formula \ref{equ1}. Let $I$ denote the set of ad impressions in the entire ad delivery period. $v(i)$ represents the valuation of the ad impression, usually indicated by the pCTR or the clicks received after the ad is displayed. $w(bid(i),\ i)$ is a winning function that measures whether the advertiser wins the auctioned impression based on its bidding strategy $bid(\cdot)$. $cost(i)$ means the cost of purchasing ad impression $i$. Apparently, the cost of buying all ad impressions should not exceed the given budget $B$. Thus, the optimization objective of the bidding strategy is transformed into solving the optimal bidding strategy $bid^*(\cdot)$ to maximize the valuation of all ad impressions purchased during the whole ad delivery period. The optimization problem with constraints can be addressed using heuristic algorithms or Lagrangian optimization methods to obtain a static optimal bidding strategy. Unfortunately, the ad impressions, the advertisers, and the bidding strategies are all variable, making the auction environment highly dynamic. Furthermore, the dynamic nature of the RTB environment leads to the static optimal bidding strategies may not work well in the new ad delivery period. We will detail the reasons in Section \ref{Representative Bidding Strategies}. \par
\subsection{RL-based Bidding strategy}
\label{RL-based Bidding strategy}
\begin{figure}[!]
	\centering
		\includegraphics[width=\textwidth]{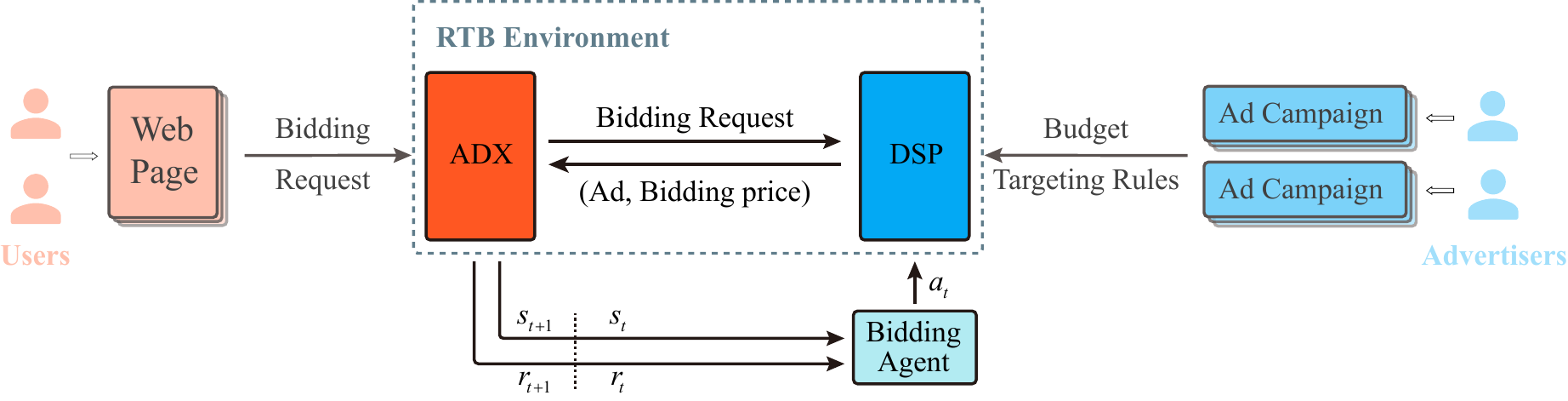}
	  \caption{An illustration of the RL bidding agent interacting with the RTB environment.}\label{fig2}
\end{figure}

The ideal bidding strategy targets a single ad impression and determines the bidding price based on its valuation and the RTB auction environment. Hence, in the latest study, the researchers solve the optimal bidding strategy by RL. In RL, the agent interacts with the environment by sequentially taking actions, observing results, and altering its behaviors to maximize the cumulative reward. Figure \ref{fig2} illustrates the interaction of the RL-based bidding agent with the RTB environment. The environment contains the ADX and DSP. The bidding agent interacts with the RTB environment by continuously bidding for sequentially arriving ad impressions, simultaneously observing the bidding results, and optimizing its strategy to maximize the ad revenue.  \par
The bidding decision process is typically modeled as an episodic MDP $<\mathcal{S},\ \mathcal{A},\ \mathcal{P},\ \mathcal{R}>$ in the RL context. An episode contains the impressions that satisfy the advertiser’s targeting rules during the ad delivery period. In RTB, the core components of MDP are usually defined as:
\begin{itemize}
\item State $\mathcal{S}$: state $s_t$ reflects the RTB auction environment when the impression arrives.
\item Action $\mathcal{A}$: action $a_t=\pi(s_t)$ is defined as the bidding price of the impression base on state $s_t$.
\item State transition dynamics $\mathcal{P}$: $p(s_{t+1}|s_t, a_t)$ denotes the probability that the agent transforms to state $s_{t+1}$ by performing action $a_t$ in state $s_t$.
\item Reward $\mathcal{R}$: $r(s_t, a_t)$ represents the immediate reward received for performing action $a_t$ in state $s_t$.
\end{itemize}
\par
The agent uses RL algorithms to optimize policy $\pi$ to maximize the cumulative reward in the ad delivery period, as shown in formula \ref{equ2}. Among three representative RL-based bidding strategies, RLB models the bidding decision by model-based RL and solves it using a dynamic programming algorithm. In contrast, DRLB and FAB model the bidding decision using model-free RL and leverage DQN \citep{DQN} and TD3 \citep{TD3} to address the optimal bidding strategy.
\begin{equation}
\begin{split}
\pi^*=\ &\underset{\pi}{\mathop{\arg \max}}\ \mathbb{E}[R|\pi] \\
where\ &R=\sum_{t=1}^T r(s_t,a_t)
\end{split}
\label{equ2}
\end{equation}
\section{Representative Bidding Strategies}
\label{Representative Bidding Strategies}
Existing bidding strategies are divided into static bidding strategies, such as LIN \citep{LIN} and ORTB \citep{ORTB}, and dynamic bidding strategies using RL, such as RLB \citep{RLB}, DRLB \citep{DRLB}, and FAB \citep{FAB}. This section describes them in detail.
\subsection{Static Bidding strategies}
\label{Static Bidding strategies}
Since RTB adopts the GSP auction mechanism, the optimal bidding price for each bidder in the GSP should be set to the valuation of the bid item to the bidder according to the optimal auction theory \citep{wang2017display}. Therefore, static linear bidding strategies based on the valuation of ad impressions are widely deployed in DSPs. The impression valuation is usually estimated using pCTR or predicted conversion rates. In practice, however, the probability of an ad conversion occurring is extremely low and always lagging, so most bidding strategies use pCTR to evaluate impressions. The general form of the linear bidding strategy is defined in formula \ref{equ3}, where the bidding price of an ad impression is generally proportional to its pCTR. $vpc$ represents the value of a click for the advertiser.
\begin{equation}
bid(i)=pCTR(i)\times vpc
\label{equ3}
\end{equation}
\par
LIN \citep{LIN} is a linear bidding strategy that uses a heuristic algorithm to find the optimal solution for the valuation of a single click based on historical data, which is used as the optimal base bid to calculate the bidding price. Formula \ref{equ4} defines the bidding function of LIN where $avg\_pCTR$ is the average pCTR of all impressions in the historical ad delivery period, and $base\_bid^*$ represents the optimal base bid. LIN sets the base bid from 1 to 300 to calculate bidding price for each impression in the historical ad delivery period. The base bid that gets the most clicks will be the optimal base bid for the new ad delivery period as a fixed parameter. LIN is simple, easy to deploy, and can achieve better performance when the new ad delivery period does not significantly change from the historical ad delivery period; thus, it is deployed on a large scale in DSPs. Since the LIN’s bidding function parameters are fixed for the new ad delivery period, LIN can experience significant performance degradation when the RTB auction environment changes. We will specify this in Section \ref{Performance of Static Bidding Strategies}.
\begin{equation}
bid_{LIN}(i)=pCTR(i)\times \frac{base\_bid^*}{avg\_pCTR}
\label{equ4}
\end{equation}
\par
Another representative static bidding strategy is ORTB \citep{ORTB}. It uses Lagrangian optimization methods to solve the optimization problem of bidding strategies with constraints. As shown in formula \ref{equ5}, it derives a nonlinear bidding function where $c$ and $\lambda$ are fitted from historical data. Similar to LIN’s shortcomings, when the RTB auction environment of the new ad delivery period differs significantly from the historical one, both $c$ and $\lambda$ do not vary with the RTB environment, and ORTB’s performance will notably decrease.
\begin{equation}
bid_{ORTB}(i)=\sqrt{\frac{c}{\lambda} pCTR(i) + c^2} - c
\label{equ5}
\end{equation}
\subsection{RL-based Bidding strategies}
\label{RL-based Bidding strategies}
The Internet environment is ever-changing, with hot events such as holidays, online shopping festivals, and new product launches leading to traffic outbreaks. These events can cause a surge in the number of ad impressions in RTB auctions, driving up the number of advertisers participating in the auctions. At the same time, the bidding strategies adopted by advertisers can change. These circumstances fundamentally contribute to the highly dynamic and uncertain nature of the RTB environment, and further results in the optimal bidding strategy based on historical data cannot work well in the new ad delivery period. For this highly dynamic environment, the ideal bidding strategy can theoretically adjust in real-time to changes in the environment. For example, if the market is competitive and the advertiser bids at low prices, resulting in insufficient impressions, the bidding price should be increased. On the other hand, if the market is uncompetitive and the bidding price is high, resulting in rapid budget spending, the bidding price should be adjusted downward. For sequential decision-making tasks requiring frequent interaction with the environment, RL is an excellent solution. \par
\subsubsection{RLB}
\label{RLB}
The first attempt to introduce RL into RTB is RLB \citep{RLB}. The authors model the bidding decisions for all ad impressions within an ad delivery period as a sequential MDP. The goal of the agent is to get as many clicks as possible in the new ad delivery period with the limited budget $B$. RLB establishes an independent state for each ad impression. When a new ad impression arrives, the agent observes the current environment as a state, including the remaining auction volume and budget of the ad delivery period and the feature vector of the current ad impression. It then selects an action from the predefined discrete action space as the bidding price for the current ad impression and participates in the bidding. If winning the bid, the environment will feed the pCTR of the ad impression to the agent as an immediate reward. \par
RLB uses dynamic programming to optimize the optimal action selection strategy; however, this model-based RL method requires a state transfer probability matrix. The resulting computational cost makes it impossible to deploy in real RTB applications. In addition, ad impressions reach DSP sequentially in real RTB scenarios, resulting in RLB failing to model the RTB auction environment completely. In fact, the information of the RTB environment is partially observable, which is more suitable to model using Constrained Markov Decision Process (CMDP) \citep{CMDP} and solve it using model-free RL methods. Note that RLB uses the pCTR of ad impression as the immediate reward. This reward setting can easily lead agents to blindly buy ad impressions to gain pCTR without considering the cost. \par
\subsubsection{DRLB}
\label{DRLB}
To avoid the problem of RLB, a model-free RL based bidding strategy DRLB \citep{DRLB} is proposed. Model-free RL agents learn optimal action selection strategies from past experiences without building a state transfer probability matrix. Therefore, it is particularly suitable for situations where the environment is partially known. On the other hand, model-free RL algorithms are hard to converge when the number of states is too large. To this end, the authors of DRLB abandon the practice of bidding directly for individual ad impressions. Instead, they divide an ad delivery period into 96 time slots, introducing a bidding factor for each time slot. The bidding price for each time slot’s ad impression is determined by the estimated ad impression valuation $pCTR(i)$ and the bidding factor $\lambda(t)$ as shown in formula \ref{equ6}. \par
DRLB attempts to generate bidding factors for each time slot using the value-based DQN algorithm. However, in DQN, the action space is limited and discrete, resulting in difficulty for the RL algorithm to learn the optimal bidding factor for each time slot. Therefore, DRLB gives up directly learning the optimal bidding factor for each time slot. Instead, it first obtains the initial bidding factor (formula \ref{equ8}) by a heuristic algorithm from historical data and then updates the bidding factor for each time slot by iteration. Specifically, at each time slot, the agent selects an action $\beta_\alpha(t)$ as an adjustment value from a predefined discrete action space, $\{-8\%, -3\%, -1\%, 0\%, 1\%, 3\%, 8\%\}$ and updates the bidding factor of current time slot using formula \ref{equ7}. Further, the agent calculates the bidding price for each impression according to the formula \ref{equ6}.
\begin{equation}
bid_{DRTB}(i,t)=\frac{pCTR(i)}{\lambda(t)}
\label{equ6}
\end{equation}
\begin{equation}
\begin{split}
\lambda(t)=&\ \lambda(t-1)\times (1+\beta_\alpha(t)) \\ 
where\ \beta_\alpha(t) \in \{-8&\%, -3\%, -1\%, 0\%, 1\%, 3\%, 8\%\}
\end{split}
\label{equ7}
\end{equation}
\begin{equation}
\lambda(0)=\frac{avg\_pCTR}{base\_bid^*}
\label{equ8}
\end{equation}
\par
Compared with model-based RL, model-free RL is more appropriate for partially observable environments. It observes new states from the environment so that avoid establishing a state transfer probability matrix, which dramatically reduces computation and storage consumption without performance loss. Nevertheless, the disadvantage of DRLB is obvious: the granularity of the adjustment value is highly dependent on the design of the action space. If the action space dimension is small, the adjustment for the bidding factor is coarse-grained, making the bidding factor obtained by the model differ from the optimal one. On the other hand, if the action space is designed too large, achieving fine-grained adjustment also brings enormous computational cost and significantly reduces the convergence speed of the learning algorithm, which aggravates the insufficient convergence of DQN itself. 
\subsubsection{FAB}
\label{FAB}
The literature \cite{FAB} presents FAB, following the time slot design of DRLB. It uses a policy-based model-free RL algorithm to model bidding decisions. The bidding function of FAB is designed in formula \ref{equ9}, where $a(t)\in [-0.99, 0.99]$ is a continuous bidding factor to adjust the bidding price. FAB leverages TD3 to learn the bidding factor for each time slot instead of the adjustment value, which allows for a more delicate and flexible adjustment of the bidding price. Moreover, FAB proposes a novel reward function. The immediate reward is derived by comparing the bidding results of LIN and FAB.
\begin{equation}
bid_{FAB}(i,t)=pCTR(i)\times \frac{base\_bid^*}{avg\_pCTR}\times \frac{1}{1+a(t)}
\label{equ9}
\end{equation}
\subsubsection{Unified Form of Bidding Function}
\label{Unified Form of Bidding Function}
In fact, both the DRLB’s and FAB’s bidding functions are transformed from LIN. Therefore, we can convert them into a unified form by a simple substitution. For DRLB, bringing formula \ref{equ8} into formula \ref{equ7} yields:
\begin{equation}
\lambda(t)=\lambda(0)\times \prod_{\tau=1}^t (1+\beta_\alpha(\tau))=\frac{acg\_pCTR}{base\_bid^*}\times \prod_{\tau=1}^t (1+\beta_\alpha(\tau))
\label{equ10}
\end{equation}
Bringing formula \ref{equ10} into formula \ref{equ6} yields:
\begin{equation}
bid_{DRLB}(i,t)=pCTR(i) \times \frac{base\_bid^*}{acg\_pCTR\times \prod_{\tau=1}^t (1+\beta_\alpha(\tau))}
\nonumber
\end{equation}
Let
\begin{equation}
base\_bid(t)=\frac{base\_bid^*}{\prod_{\tau=1}^t (1+\beta_\alpha(\tau))}
\nonumber
\end{equation}
Then we can obtain a new form of DRLB bidding function
\begin{equation}
bid_{DRLB}(i,t)=pCTR(i) \times \frac{base\_bid(t)}{avg\_pCTR}
\label{equ11}
\end{equation}
In a similar way, the bidding function of FAB can be expressed by
\begin{equation}
\begin{split}
bid_{FAB}(i,t)=pCTR(i) &\times \frac{base\_bid(t)}{avg\_pCTR} \\
where \ base\_bid(t)=&\frac{base\_bid^*}{1+a(t)}
\end{split}
\label{equ12}
\end{equation}
\par
With the formula \ref{equ11} and \ref{equ12}, DRLB and FAB introduce adjustment values related to the real-time auction environment in LIN's bidding function. Moreover, in the subsequent empirical analysis, we verify whether the adjustment of the RL-based bidding strategy is effective by observing $base\_bid(t)$’s change in each time slot.
\subsection{Discussions}
\label{Discussions}
Static bidding strategies are simple, easy to deploy, and yield decent results in scenarios with slight variation in ad delivery periods. However, they lack the ability of adaptive adjustment for the highly dynamic auction environment of RTB, so they will gradually be replaced by dynamic bidding strategies based on RL with adaptive adjustment capability. Current research on RL-based bidding strategies has achieved some desired results in academia. In addition to the results already mentioned, the literature \cite{MOTIAC} uses A3C \citep{A3C} to achieve multi-objective optimization in RTB to control ROI while maximizing revenue. The literature \cite{MARLB} and \cite{MAAB} use multi-agent RL to optimize the revenue of multiple advertisers in RTB. RL has become the ideal paradigm for bidding strategy optimization in RTB. We summarize the characteristics of the three representative RL-based bidding strategies in Table \ref{tab1}. \par
\begin{table}[!]
\small
\centering
\caption{Representative RL-based Bidding Strategies Summary}
\resizebox{\textwidth}{!}{%
\begin{tabular}{cccccc}
\toprule
\toprule
\multicolumn{1}{c}{\textbf{Strategy}} & \multicolumn{1}{c}{\textbf{Model}} & \multicolumn{1}{c}{\textbf{Algorithm}} & \multicolumn{1}{c}{\textbf{Granularity}} & \multicolumn{1}{c}{\textbf{Action}} & \multicolumn{1}{c}{\textbf{Reward}} \\ 
\midrule
RLB & Model-based & \makecell*[c]{Dynamic\\programming} & Impression & \makecell*[c]{Bidding price\\(Discrete)} & pCTR \\
DRLB & Model-free & \makecell*[c]{DQN\\(value-based)} & Time slot & \makecell*[c]{Bidding factor's\\adjustment value\\(Discrete)} & \makecell*[c]{pCTR by \\ DNN} \\
FAB & Model-free & \makecell*[c]{TD3\\(policy-based)} & Time slot & \makecell*[c]{Bidding factor\\(Continuous)} & \makecell*[c]{Manual\\definition} \\
\bottomrule
\bottomrule
\end{tabular}}%
\label{tab1}
\end{table}
As mentioned earlier, considering that the RTB environment is partially observable for agents located on DSP, we believe it is more feasible and simpler to apply a model-free RL solution for RTB from the ease of modeling. When modeling bidding decisions based on model-free RL, the ideal solution is to make fine-grained bidding for each impression; however, it is an enormous challenge for an agent to make bidding decisions for millions of ad impressions. Therefore, both DRLB and FAB divide the ad delivery period into a finite number of slots and adjust the bidding factor for each time slot to tune the bidding price according to the RTB environment. \par
\subsection{MDP modeling}
\label{MDP modeling}
For the MDP of model-free RL, there are three key components to be designed, namely, state, action, and reward function. Unfortunately, there are no theoretical studies on the design of MDP components in RTB. This subsection provides a preliminary discussion on the design of MDP components using DRLB and FAB as examples. \par
\subsubsection{State: more or less statistics?}
\label{State: more or less statistics?}
First of all, we discuss the representation of the state. In RTB, the state is the agent's observation of the environment. Since the agent is located on the DSP, it can only observe winning impressions, while information about losing impressions is missing, which makes the observation of the environment is incomplete and inaccurate. In both DRLB and FAB, the current state of the environment is represented by some key statistics from previous time slots. The state representations used for DRLB and FAB are as Table \ref{tab2}, respectively. \par
\begin{table}[!]
\centering
\caption{State representations of DRLB and FAB}
\resizebox{\textwidth}{!}{%
\begin{threeparttable}
\begin{tabular}{rl}
\toprule
\toprule
\textbf{Model} & \textbf{State representation} \\
\midrule
DRLB & \makecell*[l]{
$\bullet$ the current time step $t$ \\ 
$\bullet$ the remaining budget at time-step $B_t$\\
$\bullet$ the number of $\lambda$ adjustment opportunities left at step $t$ $ROL_t$\\
$\bullet$ the budget consumption rate $BCR_t$\\
$\bullet$ the cost per mille of impressions of the winning impressions between $t-1$ and $t$ $CPM_t$\\
$\bullet$ the auction win rate reflecting the ratio of winning impressions versus total impressions $WR_t$\\
$\bullet$ the total value of winning impressions $r_{t-1}$\\
} \\
FAB & \makecell*[l]{
$\bullet$ the average available budget ratio for the remained $T-(t-1)$ time slots $avbudget\_ratio_{t-1}$ \\ 
$\bullet$ the budget cost ratio of time slot $t-1$ $cost\_ratio_{t-1}$ \\ 
$\bullet$ the click-through rate of time slot $t-1$ $ctr_{t-1}$ \\ 
$\bullet$ the win rate of auctions of time slot $t-1$ $win\_ratio_{t-1}$\\ 
} \\
\bottomrule
\bottomrule
\end{tabular}
\begin{tablenotes}
\item Note that the symbolic definition of each state component refers to the original paper.
\end{tablenotes}
\end{threeparttable}}%
\label{tab2}
\end{table}
We can draw on both definitions when designing the state. However, we also need to consider whether these statistics are sufficient to portray the current auction environment and whether there are more useful statistics or better ways to help agents infer the current state representation of the environment. In Section \ref{Impact of State Design}, we will analyze some of these issues through empirical analysis. \par
\subsubsection{Action: discrete or continuous?}
\label{Action: discrete or continuous?}
We then discuss the action design. As discussed in Section \ref{Discussions}, ideally, each arriving impression would trigger a new state, and the action space is defined as the allowed price range, e.g. [1,300] in the RLB. However, in a real RTB system, each bidding agent faces millions of impressions every day, increasing the number of states and resulting in difficult convergence of bidding strategies. Therefore, a more appropriate solution is to define the state at the granularity of time slots. With such a state design, agents need to decide on the bidding factors for each time slot, such as DRLB and FAB. The action in the DRLB is defined as the adjustment value of the bidding factor, and it updates the bidding factor for each time slot by iteration. Moreover, in FAB, the action is defined directly as a bidding factor for each time slot. Obviously, the DRLB has a relatively narrow range of adjustment values and is a coarse-grained adjustment. The bidding performance is influenced by the adjustment value of the action space design. In addition, the bidding factor is also influenced by the initial value. If the initial value is not appropriate, subsequent adjustments are powerless and can significantly degrade performance. FAB further generates the bidding factors from a continuous space, thus making the bidding factors more consistent with the RTB environment. We will compare the effects of DRLB and FAB on the adjustment of base prices of time slots in the subsequent empirical analysis. \par
\subsubsection{Reward: from environment or manually designed, even DNN?}
\label{Reward: from environment or manually designed, even DNN?}
Finally, we discuss the impact of the immediate rewards from the environment on the optimization of the bidding strategy. RLB directly uses the pCTR of winning impressions as the immediate reward, which is inappropriate. For example, the agent may not learn to buy a small number of high-priced, high pCTR impressions with the same budget and instead go for many low-priced, low pCTR impressions because the total pCTR of these low-priced impressions is higher than the high-priced ones. However, there is almost no benefit to advertisers from these low-priced impressions, which will most likely waste this budget. \par
Actually, using pCTR as an immediate reward does not consider the budget at all, which is a critical resource in budget-constrained bidding. However, once the agent consumes more budget in the initial stages of the ad delivery period, it may miss out on potentially more valuable impressions later in the delivery period because the budget is depleted. Furthermore, because there is no punishment in the reward, this negative tendency is hardly changed by subsequent learning. DRLB likewise identifies the problem of using pCTR as an immediate reward. For this reason, DRLB designed a new reward function as shown in formula \ref{equ13}. 
\begin{equation}
r(s,a)=\max_{e\in E(s,a)} \sum^T_{t=1}r_t^{(e)}
\label{equ13}
\end{equation}
\par
DRLB believes that the return for the entire delivery period reflects how well the agent did, which is positive for all state-action pairs in that delivery period. Therefore, the DRLB takes the most significant return of all ad delivery periods, for which $(s,a)$ exists as the immediate reward for performing action $a$ in state $s$. $E(s,a)$ represents the set of ad delivery periods that existing state-action pair $(s,a)$. $r_t^{(e)}$ is the sum of the pCTR of winning impressions in time slot $t$ in ad delivery period $e$. Meanwhile, to cope with industrial-level large-scale applications, DRLB does not directly store the immediate reward of all state-action pairs but uses a deep neural network to predict it. \par
	Unlike DRLB, FAB's reward function is specially designed to compare the bidding results with LIN, giving a positive reward if the results are better and a negative reward (as a penalty) if the results are worse. This reward with an explicit penalty helps the bidding agent of FAB to converge to the optimal decision relatively quickly. Each of the above reward functions is tailored to its model. We use several reward functions to optimize the bidding performance of FAB in Section \ref{Impact on Reward Function Design} of this paper to verify the generality of several reward functions. \par
In general, current research on MDP modeling is still superficial and imperfect. Accurate MDP modeling plays a vital role in bidding strategy optimization. This will be the focus of future research on RL-based bidding strategies. \par
\section{Experiment Setting}
\label{Experiment Setting}
\subsection{Dataset}
\label{Dataset}
\subsubsection{Dataset description}
\label{Dataset description}
We conduct empirical analysis using a widely adopted benchmark dataset, iPinYou, which includes bidding logs from nine advertisers over ten ad delivery periods (2013.06.06-2013.06.15). The bidding agent does not limit the budget per delivery period for each advertiser, bidding at a fixed price of 300 (10-3 Chinese Fen) for each impression that meets the targeting rules. The dataset contains contextual information about the impression participating in the auction, such as browser, operating system, ad slot size, page type, etc. However, DSP only records the winning impression’s market price and user feedback (click).\par
Therefore, in the experiments, we only use the winning impressions for participating in the auction and do not consider those losing ones, which results in the number of impressions in the experiments being much less than the number of impressions in the actual scenario. Considering the space limitation, in this paper, we select the bidding logs of advertisers with IDs 1458, 3358, 3427, 3476 to construct four experimental datasets. The bidding logs of the first seven ad delivery periods are used as the training set, while the last three ad delivery periods build the testing set. \par
\subsubsection{Dynamics Analysis}
\label{Dynamics Analysis}
\begin{table}[!]
\centering
\caption{Dataset statistics}
\resizebox{\textwidth}{!}{%
\begin{threeparttable}
\begin{tabular}{rcccccccc}
\toprule
\toprule
\multicolumn{1}{r}{\multirow{2}[1]{*}{\textbf{Statistics}}} & \multicolumn{2}{c}{\textbf{1458}} & \multicolumn{2}{c}{\textbf{3358}} & \multicolumn{2}{c}{\textbf{3427}} & \multicolumn{2}{c}{\textbf{3476}}\\
\cmidrule{2-9}
\multicolumn{1}{c}{} & \multicolumn{1}{c}{\textbf{Training Set}} & \multicolumn{1}{c}{\textbf{Testing Set}} & \multicolumn{1}{c}{\textbf{Training Set}} & \multicolumn{1}{c}{\textbf{Testing Set}} & \multicolumn{1}{c}{\textbf{Training Set}} & \multicolumn{1}{c}{\textbf{Testing Set}} & \multicolumn{1}{c}{\textbf{Training Set}} & \multicolumn{1}{c}{\textbf{Testing Set}}\\
\midrule
\textbf{Imps} & 3083056 & 614638 & 1742104 & 300928 & 2593765 & 536795 & 1970360 & 523848 \\
\textbf{Clicks} & 2454  & 515   & 1358  & 260   & 1926  & 366 & 1027 & 287 \\
\textbf{Cost} & 212400241 & 45216454 & 160943087 & 34159766 & 210239915 & 46356518 & 156088484 & 43627585\\
\textbf{Average Cost} & 30342891 & 15072151 & 22991869 & 11386588 & 30034273 & 15452172 & 22298354 & 14542528 \\
\textbf{CTR} & 0.0796\% & 0.0838\% & 0.0780\% & 0.0864\% & 0.0743\% & 0.0682\% & 0.0521\% & 0.0548\% \\
\textbf{Average pCTR} & 0.0822\% & 0.0970\% & 0.0876\% & 0.1036\% & 0.0771\% & 0.0531\% & 0.0502\% & 0.0537\% \\
\textbf{CPM} & 68.892 & 73.565 & 92.384 & 113.514 & 81.055 & 86.357 & 79.218 & 83.282 \\
\textbf{CPC} & 86552.665 & 87798.939 & 118514.791 & 131383.715 & 109158.834 & 126657.153 & 151984.891 & 152012.491 \\
\bottomrule
\bottomrule
\end{tabular}
\begin{tablenotes}
\large
\item Average Cost indicates the average cost per day of the training or testing set. Average pCTR represents the average pCTR of ad impressions in the training or testing set. CPM means cost per thousand ad impressions. CPC denotes cost per click.
\end{tablenotes}
\end{threeparttable}}%
\label{tab3}
\end{table}
In this subsection, we analyze the dynamics of the RTB environment using dataset statistics, as shown in Table \ref{tab3}. First, the average costs per day of four testing sets shrink significantly compared to the training sets in all datasets. At the same time, the CPMs of four testing sets rise in various degrees, implying an increase in the cost of purchasing an ad impression. Changes in both statistics indicate intensified auction competition during the testing sets’ ad delivery periods. Second, on datasets 1458, 3358, and 3476, the average pCTRs of the testing sets increase. The improvement of dataset 3358 reaches 18.19\%, revealing a general improvement in the quality of ad impressions during the testing ad delivery period. \par
Similarly, there is a corresponding increase in the CTR of the testing set on three datasets, which is because the improvement in the quality of ad impressions leads to more frequent interaction between users and ads. High-quality ad impression leads to increased competition in the RTB auction environment and, therefore, a corresponding increase in CPM and CPC. Unlike other datasets, on dataset 3427, both CTR and pCTR decrease on the testing set, with a significant 31.05\% decrease in pCTR, while both CPM and CPC increase. This is because the auction environment in dataset 3427 is exceptionally competitive, which results in advertisers losing the auction for high-quality ad impressions, retreating to buy low-quality ones, and winning fewer impressions. Even if the quality of ad impressions is relatively low, their bidding prices rise due to fierce competition, causing CPM and CPC to rise. All indications from the auction results in the dataset suggest that the RTB environment is unstable and dynamic. \par
\begin{figure}[!]
	\begin{minipage}[t]{0.5\textwidth}
		\centering
		\includegraphics[scale=0.7]{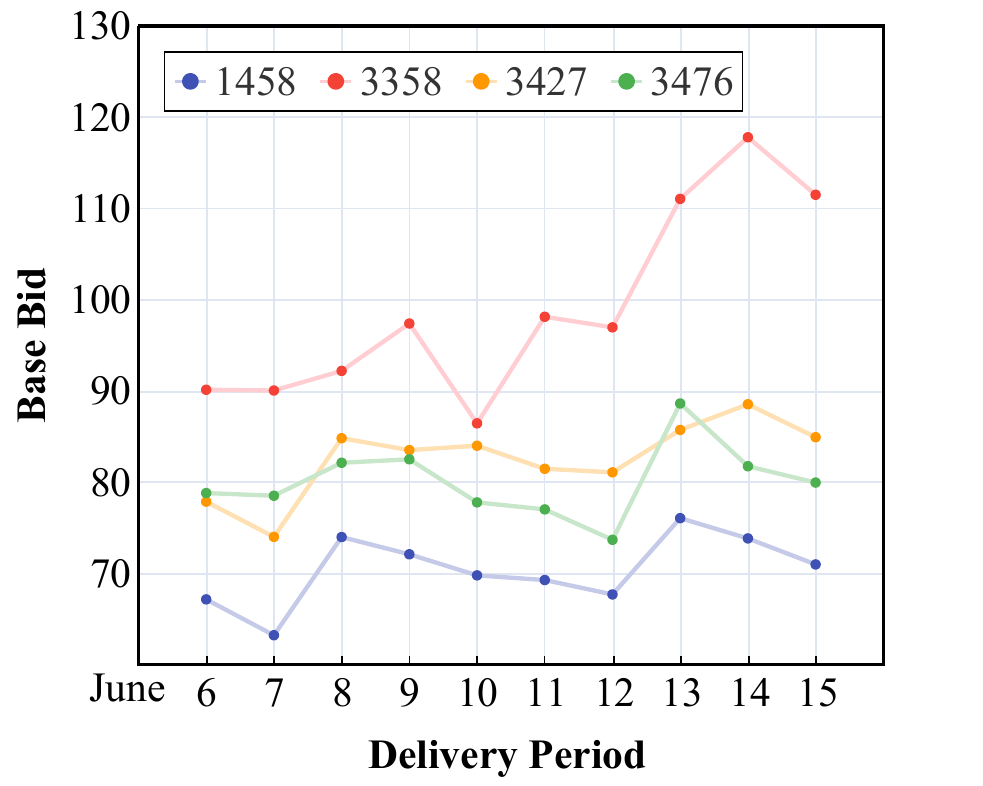}
	  	\caption{The average market price per day for training and testing sets.}		
	  	\label{fig3}
	\end{minipage}
	\hspace{0.1in}
	\begin{minipage}[t]{0.5\textwidth}
		\centering
		\includegraphics[scale=0.7]{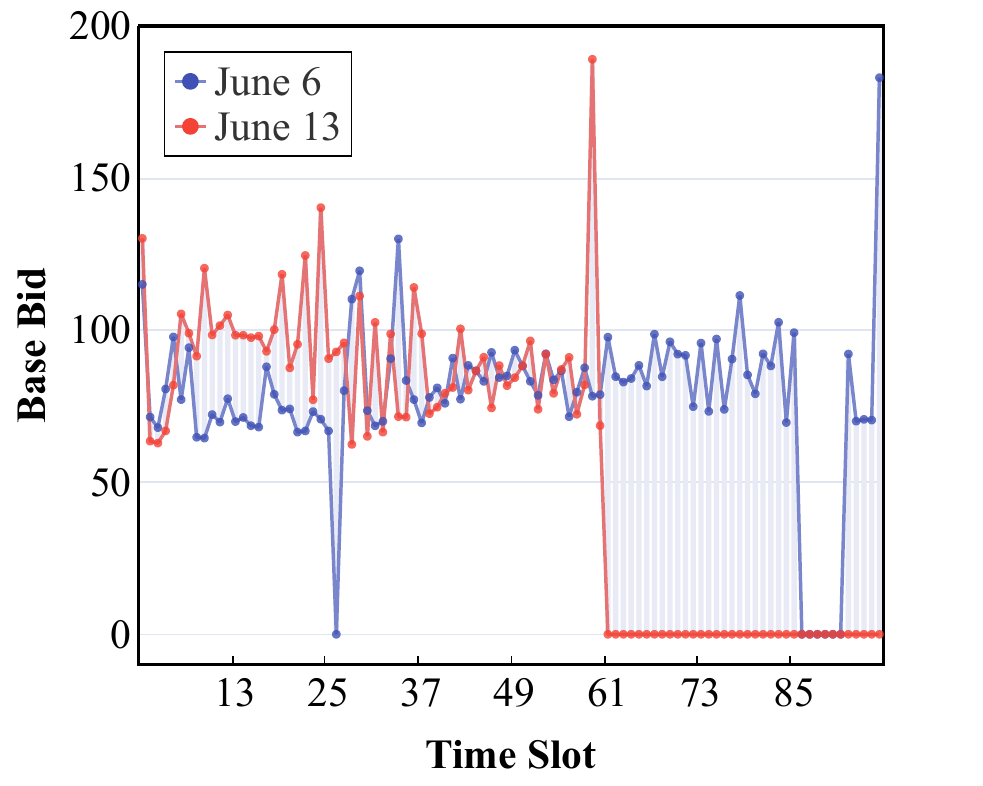}
	  	\caption{The average market price for each time slot of June 6 and June 13 in dataset 3358.}						
	  	\label{fig5}
	\end{minipage}
\end{figure}
\begin{figure}[!]
	\centering
		\includegraphics[]{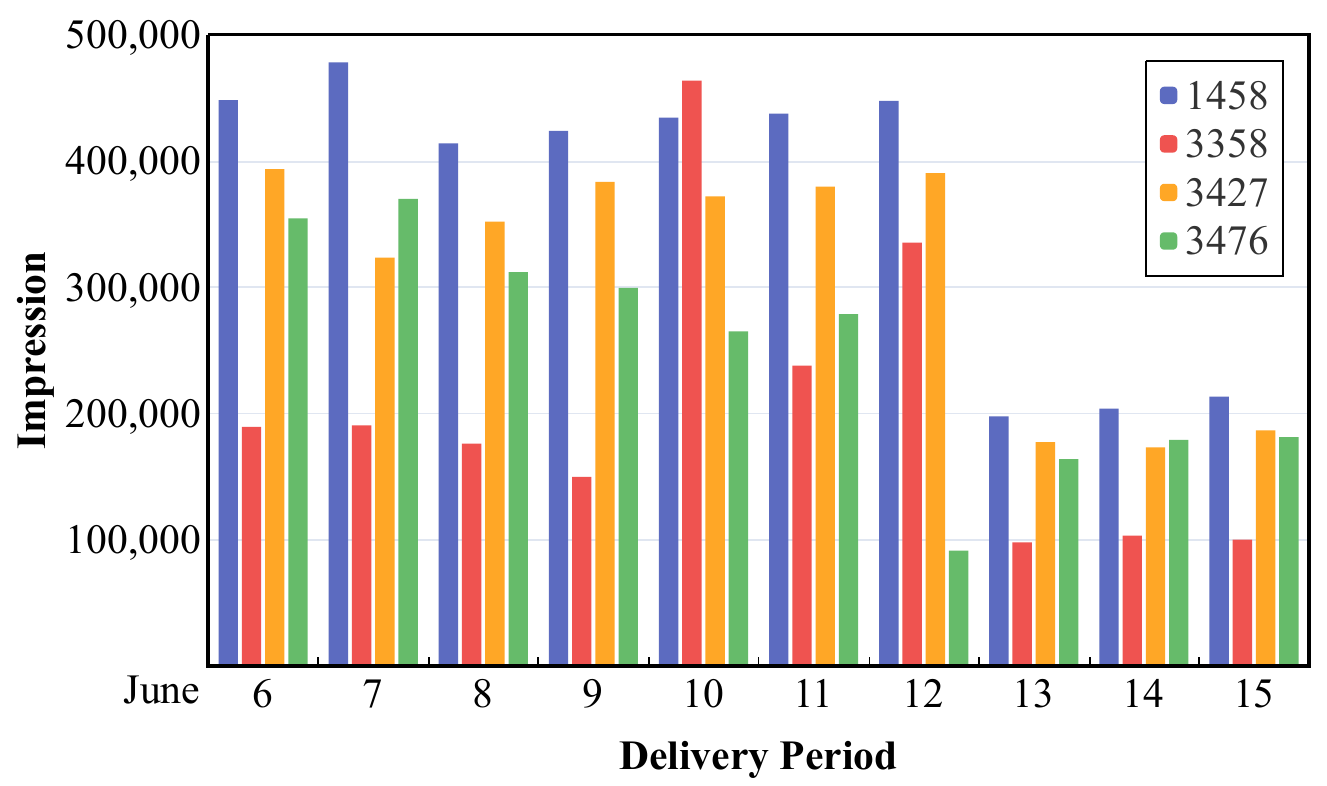}
	  \caption{The number of ad impression per day for training and testing sets.}\label{fig4}
\end{figure}

The average market price per day and the number of ad impression per day for the training and testing sets are shown in Figure \ref{fig3} and \ref{fig4} respectively. As shown in Figure \ref{fig4}, the number of ad impressions per day in the testing set in all datasets decreases sharply compared to the training set, and the average market price per day increases compared to the training set. Similar to the previous analysis results, the changes in the two statistics point to increased competition in the auction environment, reflecting the dynamic nature of the RTB environment. It is worth noting that there is a significant rise in the number of ad impressions on June 10 of the training set of dataset 3358. In the training set of dataset 3476, the number of ad impressions on June 12 significantly drops from the previous day. This phenomenon indicates that the dynamic nature of the RTB auction environment exists not only between the training and testing sets but also varies from day to day. \par
Further, we list the average market price for each time slot of June 6 and June 13 in dataset 3358, the first day of the training set and the testing set, respectively, as shown in Figure \ref{fig5}. The average market price fluctuates significantly during the day. For example, between time slot 37 and 57, the average market price remains stable over the two days, while there are large fluctuations between slot 8 and 27. Note that advertiser 3358 did not get ad impressions from the 60th time slot on June 13, and thus the corresponding average market price became 0. This particular situation may never occur in the training set. The above results show that the RTB auction environment is also highly dynamic during the day, which requires the bidding strategy to allocate the budget to each time slot rationally and improve the budget efficiency. \par
\subsection{Setup}
\label{Setup}
In this subsection, we describe the settings of the budget, the CTR estimator, and the metrics in the experiments. \par
\subsubsection{Budget Setting}
\label{Budget Setting}
The budget largely determines the upper limit of the performance of the bidding strategy. In the iPinYou dataset, the market price of each ad impression is recorded, which is the cost to the advertiser to purchase the impression. We can use the sum of the market price of all ad impressions for each day as the budget for the day. However, in this case, the bidding strategy only needs to bid 300 to buy all ad impressions to get the best bidding performance. Therefore, we use 1/2, 1/4, 1/8, and 1/16 of the total daily cost as the day's budget to measure the bidding strategy's sensitivity to budget changes. Note that the day's budget can only be used on the same day. \par
\subsubsection{CTR Estimator}
\label{CTR Estimator}
The bidding strategies covered in this paper are based on the estimated valuation (pCTR) of an ad impression \citep{richardson2007predicting}. The bidding agent first calculates the pCTR of the ad impression and derives the corresponding bidding price. So, we train a CTR estimator for each advertiser using FM \citep{FM} to calculate the pCTR of each ad impression. Table \ref{tab4} shows the AUC for each estimator. It can be observed that there are some variations in the AUC of different datasets. The CTR estimator achieves good performance on 1458, 3358, and 3427. Besides, the AUC of 3476 is much lower than the other three datasets, mainly due to the very sparse click behavior of dataset 3476. In our experiments, we use the four CTR estimators to evaluate the valuation of ad impression. \par
\begin{table}[t]
\small
\centering
\caption{The AUC of four CTR estimators}
\begin{tabular}{ccccc}
\toprule
\toprule
\textbf{Dataset} & 1458 & 3358 & 3427 & 3476 \\
\midrule
\textbf{AUC} & 0.8365 & 0.8892 & 0.8521 & 0.7026\\
\bottomrule
\bottomrule
\end{tabular}
\label{tab4}
\end{table}
\subsubsection{Metrics}
\label{Metrics}
The goal of the bidding strategy is to optimize advertisers' revenue within a constrained budget. Typically, we compare the number of clicks on ad impressions earned by different bidding strategies. In addition, some of the RL-based bidding strategies use pCTR as the immediate reward, so we also include the total pCTR of winning impressions in the comparison. Finally, to help researchers quickly reproduce the experiments in this paper, we have fully open-sourced all the experimental code on \href{https://github.com/hzn666/RLBid\_EA}{Github}. \par
\section{Empirical Analysis}
\label{Empirical Analysis}
In this section, we first compare the performance of two static bidding strategies and discuss their problems. Then, we compare three RL-based bidding strategies in detail. In particular, we discuss the impacts of three components on the bidding performance in the model-free RL framework for MDP modeling. Finally, we give valuable suggestions for optimizing bidding strategies using RL based on the empirical analysis. \par
\subsection{Performance of Static Bidding Strategies}
\label{Performance of Static Bidding Strategies}
In this subsection, we evaluate the performance of two representative static bidding strategies, respectively LIN and ORTB. LIN uses a heuristic algorithm to find the optimal base bid that maximizes the total number of clicks on the training set and then applies it to bid on the three testing delivery periods. When the bidding price is greater than or equal to the recorded market price, we judge the advertiser wins the impression. The total number of clicks and total pCTR obtained during the testing periods are shown in Table \ref{tab5}. \par
\begin{table}[!]
\centering
\caption{Bidding Results of LIN}
\resizebox{\textwidth}{!}{%
\begin{tabular}{ccccccccc}
\toprule
\toprule
\multicolumn{1}{c}{\multirow{2}[1]{*}{\textbf{Budget}}} & \multicolumn{2}{c}{\textbf{1458}} & \multicolumn{2}{c}{\textbf{3358}} & \multicolumn{2}{c}{\textbf{3427}} & \multicolumn{2}{c}{\textbf{3476}}\\
\cmidrule{2-9}
\multicolumn{1}{c}{} & \multicolumn{1}{c}{\textbf{clicks}} & \multicolumn{1}{c}{\textbf{pCTR}} & \multicolumn{1}{c}{\textbf{clicks}} & \multicolumn{1}{c}{\textbf{pCTR}} & \multicolumn{1}{c}{\textbf{clicks}} & \multicolumn{1}{c}{\textbf{pCTR}} & \multicolumn{1}{c}{\textbf{clicks}} & \multicolumn{1}{c}{\textbf{pCTR}}\\
\midrule
\textbf{1/2} & 438 & 528.7304 & 222 & 267.8814 & 289 & 235.5159 & 206 & 234.8283 \\
\textbf{1/4} & 358 & 441.3804 & 196 & 227.0396 & 236 & 186.1348 & 132 & 181.9132 \\
\textbf{1/8} & 306 & 358.3400 & 178 & 200.6596 & 187 & 150.1995 & 88 & 132.6280 \\
\textbf{1/16} & 261 & 291.3947 & 167 & 177.9763 & 169 & 125.5816 & 57 & 93.5930 \\
\bottomrule
\bottomrule
\end{tabular}}%
\label{tab5}
\end{table}
Combining Table \ref{tab3} and Table \ref{tab5}, we can observe a gap between clicks obtained under different budget conditions and the actual clicks. As shown in formula \ref{equ4}, the bidding price of an ad impression depends on the ad impression valuation and is also closely related to the base bid. The optimal base bid in LIN is derived from historical delivery periods. Hence, LIN achieves poor performance when the RTB environment for the new ad delivery period deviates from the historical one. Notably, when the budget is set to 1/16 of the total cost, LIN experienced an early stop on dataset 1458. The early stop happens when the budget is spent out early before the end of the delivery period, making it impossible to bid on all subsequent ad impressions. Therefore, the advertiser wants to avoid the early stop by adjusting its budget spending rate. \par
Further, we analyze the impact of environmental dynamics on LIN’s performance. First, we assume that all ad impressions on the testing set are known and execute a heuristic algorithm to get the optimal base bid of the testing set and bid based on that optimal base bid. This method yields more clicks because the base bid is directly based on the testing set and is more in line with the current RTB environment. \par
\begin{table}[!]
\centering
\caption{The optimal base bid and its bidding results for the training and testing sets}
\resizebox{\textwidth}{!}{%
\begin{threeparttable}
\begin{tabular}{cccccccccc}
\toprule
\toprule
\multicolumn{2}{c}{\multirow{2}[1]{*}{\textbf{Budget}}} & \multicolumn{2}{c}{\textbf{1458}} & \multicolumn{2}{c}{\textbf{3358}} & \multicolumn{2}{c}{\textbf{3427}} & \multicolumn{2}{c}{\textbf{3476}}\\
\cmidrule{3-10}
\multicolumn{2}{c}{} & \multicolumn{1}{c}{\textbf{base bid}} & \multicolumn{1}{c}{\textbf{clicks}} & \multicolumn{1}{c}{\textbf{base bid}} & \multicolumn{1}{c}{\textbf{clicks}} & \multicolumn{1}{c}{\textbf{base bid}} & \multicolumn{1}{c}{\textbf{clicks}} & \multicolumn{1}{c}{\textbf{base bid}} & \multicolumn{1}{c}{\textbf{clicks}}\\
\midrule
\multirow{2}[0]{*}{\textbf{1/2}} & \textbf{train} & 147 & 438 & 185 & 222 & 198 & 289 & 126 & 206 \\
& \textbf{test} & 186 & 442 & 297 & 230 & 238 & 326 & 162 & 226 \\
\midrule
\multirow{2}[0]{*}{\textbf{1/4}} & \textbf{train} & 72 & 358 & 89 & 196 & 99 & 236 & 62 & 132 \\
& \textbf{test} & 83 & 359 & 135 & 206 & 103 & 265 & 81 & 154 \\
\midrule
\multirow{2}[0]{*}{\textbf{1/8}} & \textbf{train} & 44 & 306 & 58 & 178 & 60 & 187 & 39 & 88 \\
& \textbf{test} & 52 & 306 & 75 & 181 & 59 & 224 & 53 & 98 \\
\midrule
\multirow{2}[0]{*}{\textbf{1/16}} & \textbf{train} & 30 & 261 & 40 & 167 & 38 & 169 & 27 & 57 \\
& \textbf{test} & 34 & 263 & 46 & 168 & 38 & 185 & 33 & 61 \\
\bottomrule
\bottomrule
\end{tabular}
\begin{tablenotes}
\footnotesize
\item The train row gives the optimal base bid derived from the training set and its clicks on the testing set; The test row gives the optimal base bid derived from the testing set and its clicks on the testing set.
\end{tablenotes}
\end{threeparttable}}%
\label{tab6}
\end{table}
The results are shown in Table \ref{tab6}. There is a significant difference between the optimal base bids of the training set and the testing set. Only on dataset 1458 and with the budget set to 1/8 the optimal base bid of the training set gets the same clicks as the optimal base bid of the testing set. However, in other experimental settings, the optimal base bid of the testing set gets a higher number of clicks than the optimal base bid of the training set. We introduce formula \ref{equ14} to quantitatively describe the deviation of the optimal base bid of the training and testing sets. $b_i$ represents the optimal base bid of the testing set and the optimal base bids of each testing day, respectively. $b_0$ represents the optimal base bid of the training set. 
\begin{equation}
Deviation=\frac{\sqrt{\sum_{i=4}^4(\frac{b_i}{b_0}-1)^2}}{4}
\label{equ14}
\end{equation}
\par
The deviations under the four budget conditions in the four datasets are given in Table \ref{tab7}. On the one hand, all ad campaigns have a high degree of variability in the training and testing sets, which is caused by the instability of the auction environment. On the other hand, 3358 is significantly more dynamic than the other datasets, driven by the different industries in which the ad campaigns are located. \par
\begin{table}[!]
\small
\centering
\caption{Deviations under the four budget conditions in the four datasets}
\begin{tabular}{ccccc}
\toprule
\toprule
\textbf{Budget} & \textbf{1458} & \textbf{3358} & \textbf{3427}& \textbf{3476} \\
\midrule
\textbf{1/2} & 25.83\% & 57.38\% & 15.65\% & 27.63\% \\
\textbf{1/4} & 24.24\% & 79.77\% & 7.92\% & 29.10\% \\
\textbf{1/8} & 19.75\% & 50.18\% & 7.31\% & 27.52\% \\
\textbf{1/16} & 21.47\% & 38.12\% & 1.32\% & 18.33\% \\
\bottomrule
\bottomrule
\end{tabular}
\label{tab7}
\end{table}
The bidding performance of ORTB under four budget conditions in four datasets is presented in Table \ref{tab8}, which the $c$ and $\lambda$ in formula \ref{equ5} are taken as 34 and $5.2\times 10^{-7}$. ORTB outperforms LIN in some cases. However, due to the inherent flaw of the static bidding strategy, which cannot perceive the dynamic nature of the RTB environment, ORTB performs even worse than LIN on dataset 1458. \par
\begin{table}[!]
\centering
\caption{Bidding Results of ORTB}
\resizebox{\textwidth}{!}{%
\begin{tabular}{ccccccccc}
\toprule
\toprule
\multicolumn{1}{c}{\multirow{2}[1]{*}{\textbf{Budget}}} & \multicolumn{2}{c}{\textbf{1458}} & \multicolumn{2}{c}{\textbf{3358}} & \multicolumn{2}{c}{\textbf{3427}} & \multicolumn{2}{c}{\textbf{3476}}\\
\cmidrule{2-9}
\multicolumn{1}{c}{} & \multicolumn{1}{c}{\textbf{clicks}} & \multicolumn{1}{c}{\textbf{pCTR}} & \multicolumn{1}{c}{\textbf{clicks}} & \multicolumn{1}{c}{\textbf{pCTR}} & \multicolumn{1}{c}{\textbf{clicks}} & \multicolumn{1}{c}{\textbf{pCTR}} & \multicolumn{1}{c}{\textbf{clicks}} & \multicolumn{1}{c}{\textbf{pCTR}}\\
\midrule
\textbf{1/2} & 383 & 470.4769 & 224 & 273.5141 & 300 & 244.4029 & 215 & 241.0612 \\
\textbf{1/4} & 308 & 386.0084 & 210 & 247.5986 & 250 & 199.0784 & 142 & 181.1455 \\
\textbf{1/8} & 234 & 294.7192 & 182 & 212.2620 & 198 & 161.3886 & 85 & 129.8643 \\
\textbf{1/16} & 200 & 247.9527 & 174 & 184.5505 & 167 & 130.1005 & 61 & 102.1966 \\
\bottomrule
\bottomrule
\end{tabular}}%
\label{tab8}
\end{table}
In summary, we verify the performance of the two static bidding strategies separately. The experimental results demonstrate that for the highly dynamic RTB environment, the static bidding strategy is indeed challenging to achieve the optimal performance because it cannot dynamically adjust the bidding function according to the environment. \par
\subsection{Performance of RL-based Bidding Strategies}
\label{Performance of RL-based Bidding Strategies}
In this set of experiments, we focus on evaluating the performance of three RL-based bidding strategies. The experimental results are listed in Table \ref{tab9}, and we also present the results of LIN and ORTB for comparison. The optimal results occur in the RL-based bidding strategy under four budget conditions in all four datasets, both in clicks and pCTR, verifying the effectiveness of the RL-based bidding strategy. This is mainly because the static strategies cannot adapt to the dynamically changing RTB environment, resulting in poorer results. The RL-based bidding strategy can dynamically adjust the bids for ad impressions based on the auction environment and budget, showing excellent performance. \par
\begin{table}[!]
\centering
\caption{Bidding Results of RL-based Bidding Strategies}
\resizebox{\textwidth}{!}{%
\begin{tabular}{cccccccccc}
\toprule
\toprule
\multicolumn{2}{c}{\multirow{2}[1]{*}{\textbf{Budget}}} & \multicolumn{2}{c}{\textbf{1458}} & \multicolumn{2}{c}{\textbf{3358}} & \multicolumn{2}{c}{\textbf{3427}} & \multicolumn{2}{c}{\textbf{3476}}\\
\cmidrule{3-10}
\multicolumn{2}{c}{} & \multicolumn{1}{c}{\textbf{clicks}} & \multicolumn{1}{c}{\textbf{pCTR}} & \multicolumn{1}{c}{\textbf{clicks}} & \multicolumn{1}{c}{\textbf{pCTR}} & \multicolumn{1}{c}{\textbf{clicks}} & \multicolumn{1}{c}{\textbf{pCTR}} & \multicolumn{1}{c}{\textbf{clicks}} & \multicolumn{1}{c}{\textbf{pCTR}}\\
\midrule
\multirow{5}[0]{*}{\textbf{1/2}} & \textbf{LIN} & 438 & 528.7304 & 222 & 267.8814 & 289 & 235.5159 & 206 & 234.8283 \\
& \textbf{ORTB} & 383 & 470.4769 & 224 & 273.5141 & 300 & 244.4029 & 215 & 241.0612 \\
& \textbf{RLB} & 424 & 508.8491 & 227 & 270.8131 & 295 & 235.1985 & 203 & 232.8529 \\
& \textbf{DRLB} & 442 & 527.0697 & \textbf{233} & \textbf{279.8474} & \textbf{325} & \textbf{254.7858} & 221 & 242.7533 \\
& \textbf{FAB} & \textbf{453} & \textbf{533.2005} & 232 & 279.6285 & 320 & 252.1587 & \textbf{229} & \textbf{244.2495} \\
\midrule
\multirow{5}[0]{*}{\textbf{1/4}} & \textbf{LIN} & 358 & 441.3804 & 196 & 227.0396 & 236 & 186.1348 & 132 & 181.9132 \\
& \textbf{ORTB} & 308 & 386.0084 & 210 & 247.5986 & 250 & 199.0784 & 142 & 181.1455 \\
& \textbf{RLB} & 350 & 419.6544 & 200 & 231.4942 & 249 & 192.4976 & 146 & 183.0156 \\
& \textbf{DRLB} & 361 & 439.2041 & \textbf{208} & \textbf{241.9180} & 267 & 209.3853 & 148 & \textbf{192.2448} \\
& \textbf{FAB} & \textbf{373} & \textbf{443.0800} & 207 & 241.4922 & \textbf{275} & \textbf{212.4110} & \textbf{156} & 184.7556 \\
\midrule
\multirow{5}[0]{*}{\textbf{1/8}} & \textbf{LIN} & 306 & 358.3400 & 178 & 200.6596 & 187 & 150.1995 & 88 & 132.6280 \\
& \textbf{ORTB} & 234 & 294.7192 & 182 & 212.2620 & 198 & 161.3886 & 85 & 129.8643 \\
& \textbf{RLB} & 289 & 343.9998 & 170 & 192.6070 & 205 & 161.1381 & 90 & 127.4535 \\
& \textbf{DRLB} & 306 & 360.9919 & 184 & 210.4508 & 223 & \textbf{175.0740} & 100 & \textbf{143.4913} \\
& \textbf{FAB} & \textbf{314} & \textbf{366.3085} & \textbf{189} & \textbf{212.4545} & \textbf{229} & 174.5927 & \textbf{101} & 134.2651 \\
\midrule
\multirow{5}[0]{*}{\textbf{1/16}} & \textbf{LIN} & 261 & 291.3947 & 167 & 177.9763 & 169 & 125.5816 & 57 & 93.5930 \\
& \textbf{ORTB} & 200 & 247.9527 & 174 & 184.5505 & 167 & 130.1005 & 61 & 102.1966 \\
& \textbf{RLB} & 253 & 284.1054 & 141 & 153.4800 & 179 & 136.3995 & 63 & 90.6204 \\
& \textbf{DRLB} & 270 & 304.7552 & \textbf{172} & \textbf{183.5584} & 186 & \textbf{144.2147} & 64 & 97.6938 \\
& \textbf{FAB} & \textbf{273} & \textbf{307.1864} & \textbf{172} & 181.3249 & \textbf{195} & 142.5405 & \textbf{67} & \textbf{101.9895} \\
\bottomrule
\bottomrule
\end{tabular}}%
\label{tab9}
\end{table}
Comparing the three RL-based bidding strategies, among the 16 settings, FAB gets the most clicks in 13 settings and the most pCTRs in 8 settings, and DRLB gets the most clicks and pCTRs in 4 and 8 settings, respectively. RLB performs the worst, even inferior to LIN in 7 settings. RLB can allocate budgets to the entire ad delivery period and achieve bids at the granularity of a single ad impression. However, its bids for ad impressions are searched on a lookup table built from historical data, making RLB insensitive to environmental changes. DRLB and FAB achieve good results over RLB. While RLB considers optimizing the bids for a single ad impression, DRLB and FAB choose to divide the ad delivery period into multiple time slots and adjust the bids for ad impressions on a time slot basis. Such schemes can significantly reduce the number of states in the RL environment and enhance the model's ability to perceive changes in the environment. \par
DRLB's performance is moderate, primarily because DRLB utilizes DQN to optimize the bidding strategy, which can only adjust the bidding factor by selecting an adjustment value from a manually discrete action space. This coarse-grained adjustment is, in most cases, very far from the optimal one, which is why DRLB performance is weaker than FAB. Also, DRLB uses the pCTR of the ad impression as the immediate reward, so it gets the most pCTR in 8 settings. However, using pCTR as the immediate reward is inappropriate because it induces the bidding agent to buy too many ad impressions blindly without considering the cost, resulting in budget spending out, i.e., early stop. In addition, there is no negative value for the pCTR reward, leaving no explicit penalty for any agent's action, which could weaken the agent's performance. \par
FAB performs the best of all RL-based bidding strategies. FAB follows DRLB's time slot design and directly generates the bidding factor for each time slot, adjusting the bids of ad impressions during the time slot. At the same time, the action space of FAB is continuous, and it enables fine-grained bidding price adjustment. In addition, FAB designs a new reward function that derives an immediate reward for the current time slot by comparing the bidding results of the time slot with LIN. This reward function abandons the traditional idea of using pCTR as the immediate reward. It adds a negative reward to penalize the agent's poor performance, enabling the agent to improve bidding performance based on LIN. Besides, FAB directly generates bidding factors for each time slot rather than adjustment values of the bidding factors. The learning objective of the agent is closer to the actual optimization objective. \par
Overall, this subsection highlights a comparison of three representative RL-based bidding strategies. Experimental results reveal that the model-based RL method is ineffective in optimizing the bidding strategy for the highly dynamic RTB environment because it is challenging to model the environment accurately. Using the model-free RL algorithm is a better choice. Further, FAB generates the bidding factor for each time slot from the continuous action space, which is the referenceable solution. \par
\subsection{Impacts of MDP components on bidding performance}
\label{Impacts of MDP components on bidding performance}
The empirical analysis above shows that model-free RL is a promising solution to optimizing the bidding strategy in RTB. To this end, this subsection further analyzes the impact of three critical MDP components of model-free RL-based bidding strategies on bidding performance from an experimental perspective. Specifically, we discuss three aspects of state, action, and reward function design. \par
\subsubsection{Impact of State Design}
\label{Impact of State Design}
In RL, the agent needs to observe the real world, derives a valid representation of the environment as the state, and then serve as the basis for action selection. Therefore, state design is essential. In RTB, the bidding decision is related to the valuation of the impression and the RTB auction environment, and the advertiser's current available budget. Therefore, both DRLB and FAB use the statistical information of the latest finished time slot to define the current state, as defined in Section \ref{Discussions}. For example, they both use the budget cost rate to indicate the budget spend speed. Win rate, CPM, and CTR metrics represent the effectiveness of the current bidding strategy and the level of competition in the market. In fact, since the agent is located on the DSP, it is only partially aware of the environment as informed by the ADX, making its observation of the environment inadequate. Representing the current state through statistical information is a feasible approach, but there is still much room for improvement. Which metrics positively affect state representation and which metrics appear to be helpful but potentially could introduce much confounding noise need to be further explored by researchers. \par
\begin{table}[t]
\small
\centering
\caption{State design schemes}
\begin{tabular}{rl}
\toprule
\toprule
\textbf{Scheme} & \textbf{Definition} \\
\midrule
$State_6$ & $B_t$, $ROL_t$, $BCR_t$, $CPM_t$, $WR_t$, $r_{t-1}$ \\
$State_5$ & $B_t$, $BCR_t$, $CPM_t$, $WR_t$, $r_{t-1}$ \\
$State_4$ & $B_t$, $BCR_t$, $CPM_t$, $WR_t$ \\
$State_3$ & $BCR_t$, $CPM_t$, $WR_t$ \\
$State_2$ & $BCR_t$, $CPM_t$ \\
$State_1$ & $BCR_t$ \\
\bottomrule
\bottomrule
\end{tabular}
\label{tab10}
\end{table}
In this group of experiments, we conduct a preliminary attempt to analyze which metrics are valued based on dataset 1458 in DRLB. We rank the statistical indicators used to define the state in the DRLB in order of importance, discard the statistical indicators we consider least important one by one and use the remaining ones to define the state. The experimental scheme is designed as Table \ref{tab10}, and the corresponding bidding results are shown in Table \ref{tab11}. The results show that as the number of statistical indicators gradually decreases, the performance of the bidding strategy does not decrease significantly but shows a slight trend of first increasing and then decreasing. The more statistical indicators employed are not better, as some do not portray the state obviously, but may introduce noise that affects the agent's perception of the environment. In addition, we find that the solution with the three metrics of budget consumption rate, CPM, and win rate in the state definition obtains better results, indicating that these three metrics are more valuable for portraying the RTB auction environment. \par
\begin{table}[!]
\centering
\caption{Bidding results of different state design}
\resizebox{\textwidth}{!}{%
\begin{tabular}{ccccccccc}
\toprule
\toprule
\multicolumn{1}{c}{\multirow{2}[1]{*}{\makecell*[c]{\textbf{State}\\ \textbf{Design}}}} & \multicolumn{2}{c}{\textbf{1458}} & \multicolumn{2}{c}{\textbf{3358}} & \multicolumn{2}{c}{\textbf{3427}} & \multicolumn{2}{c}{\textbf{3476}}\\
\cmidrule{2-9}
\multicolumn{1}{c}{} & \multicolumn{1}{c}{\textbf{clicks}} & \multicolumn{1}{c}{\textbf{pCTR}} & \multicolumn{1}{c}{\textbf{clicks}} & \multicolumn{1}{c}{\textbf{pCTR}} & \multicolumn{1}{c}{\textbf{clicks}} & \multicolumn{1}{c}{\textbf{pCTR}} & \multicolumn{1}{c}{\textbf{clicks}} & \multicolumn{1}{c}{\textbf{pCTR}}\\
\midrule
$State_6$ & 442 & 525.3456 & 362 & 440.3299 & 308 & 359.7897 & 266 & 300.4850 \\
$State_5$ & 445 & 531.0983 & 365 & 439.4166 & 308 & 362.2590 & 266 & 298.3722 \\
$State_4$ & 449 & 528.4864 & 359 & 433.8279 & 307 & 357.5095 & 265 & 297.4223 \\
$State_3$ & 446 & 532.9431 & 361 & 437.0696 & 308 & 363.5619 & 272 & 301.3293 \\
$State_2$ & 446 & 531.1273 & 365 & 445.9889 & 308 & 360.2119 & 270 & 297.4041 \\
$State_1$ & 439 & 524.6575 & 360 & 442.1739 & 308 & 360.2210 & 265 & 299.1287 \\
\bottomrule
\bottomrule
\end{tabular}}%
\label{tab11}
\end{table}
In addition, both DRLB and FAB divide the ad delivery period into multiple time slots and adjust or generate bidding factors of each time slot to regulate the bids for the ad impression. DRLB divides the ad delivery period into 96 time slots, compared to 24 for FAB. The way the ad delivery period is divided determines the number of states, which is a potential factor in bidding performance. To this end, we conducted experiments using DRLB and FAB on dataset 1458, dividing the ad delivery periods by 24, 48, and 96 time slots, respectively, to compare the impact of the different number of time slots on bidding performance. The results are shown in Table \ref{tab12}. \par
\begin{table}[!]
\centering
\caption{Bidding results of different number of time slots}
\resizebox{\textwidth}{!}{%
\begin{tabular}{cccccccccc}
\toprule
\toprule
\multicolumn{2}{c}{\multirow{2}[1]{*}{\textbf{Model}}} & \multicolumn{2}{c}{\textbf{1/2}} & \multicolumn{2}{c}{\textbf{1/4}} & \multicolumn{2}{c}{\textbf{1/8}} & \multicolumn{2}{c}{\textbf{1/16}}\\
\cmidrule{3-10}
\multicolumn{2}{c}{} & \multicolumn{1}{c}{\textbf{clicks}} & \multicolumn{1}{c}{\textbf{pCTR}} & \multicolumn{1}{c}{\textbf{clicks}} & \multicolumn{1}{c}{\textbf{pCTR}} & \multicolumn{1}{c}{\textbf{clicks}} & \multicolumn{1}{c}{\textbf{pCTR}} & \multicolumn{1}{c}{\textbf{clicks}} & \multicolumn{1}{c}{\textbf{pCTR}}\\
\midrule
\multirow{3}[0]{*}{\textbf{DRLB}} & 24 & 444 & 523.8169 & 365 & 447.1079 & 308 & 359.8229 & 269 & 304.0549 \\
& 48 & 441 & 530.1044 & 363 & 439.3007 & 308 & 360.1077 & 265 & 296.8528 \\
& 96 & 442 & 527.0697 & 361 & 439.2041 & 306 & 362.7918 & 270 & 304.7552 \\
\midrule
\multirow{3}[0]{*}{\textbf{FAB}} & 24 & 452 & 532.5847 & 373 & 438.0073 & 315 & 365.0421 & 276 & 305.0419 \\
& 48 & 453 & 530.9745 & 372 & 443.8402 & 312 & 363.1831 & 274 & 302.2823 \\
& 96 & 453 & 533.2005 & 373 & 443.0800 & 314 & 366.3085 & 273 & 307.1864 \\
\bottomrule
\bottomrule
\end{tabular}}%
\label{tab12}
\end{table}
As can be seen from the table, the number of time slots does not significantly involve bidding performance. The finer the time slot division, the greater the number of states in the MDP, and the finer the adjustment of the base bid can be. However, the time interval between the two states is also shorter. For agents that can only represent the state through statistics of time slots, it may be difficult to distinguish differences in successive states and select differentiated actions. So, the finer time slot division in this experiment does not bring performance improvement. \par
\subsubsection{Impact of Action Design}
\label{Impact on Action Design}
In this subsection, we experimentally analyze the impact of the action design on the performance of DRLB and FAB. Specifically, DRLB employs a discrete action space that updates the bidding factors for each time slot by iterative adjustment. FAB defines a continuous action space of a given range, directly generating bidding factors for each time slot. Although there are differences in the design of the DRLB and FAB actions, both the DRLB's and FAB's bidding functions can be converted to a form using the base bid of each time slot by a simple substitution, as formula \ref{equ11} and \ref{equ12}. Therefore, in this set of experiments, we compare the base bids of DRLB and FAB in each time slot to examine whether the action selection or generation of the bidding agent achieves the effect. \par
\begin{figure}[!]
	\centering
		\includegraphics[]{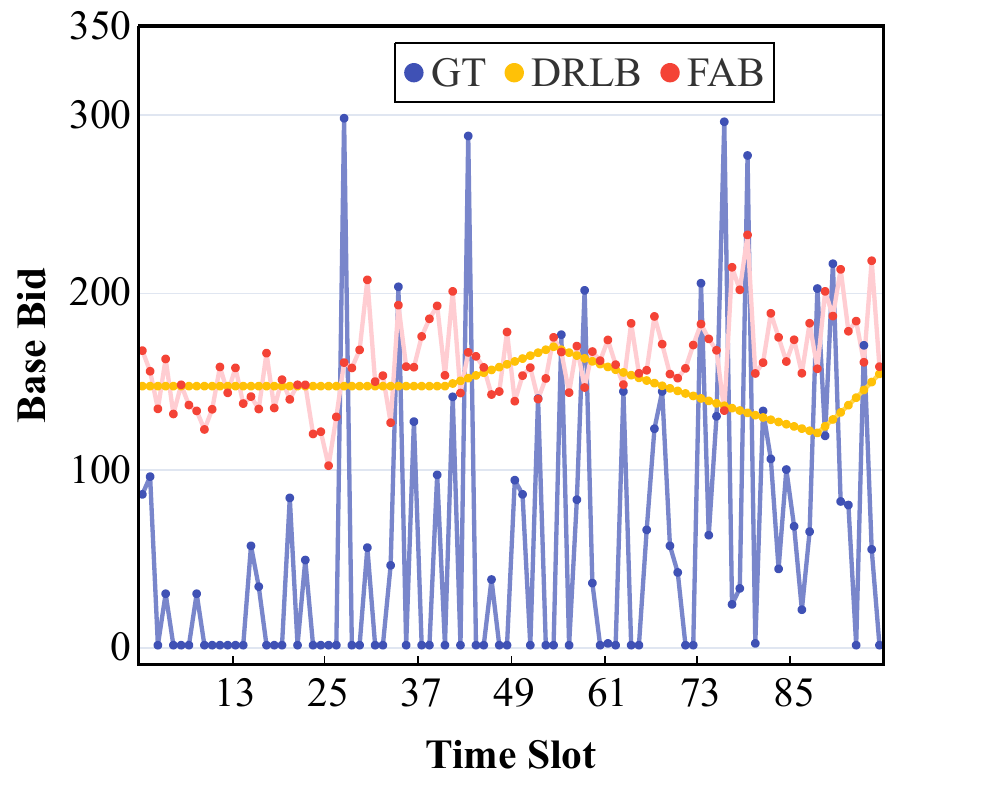}
	  \caption{The Base bid for each time slot of the three bidding strategies.}\label{fig6}
\end{figure}

We compare the base bids of the two bidding strategies for each time slot on day 3 (June 15) of the testing set in dataset 1458 under 96 time slots with a budget condition of 1/2. Meanwhile, we assume that the testing set is known and use the heuristic algorithm to calculate the optimal base bid for each time slot separately as the ground truth (GT) for comparison. The budget for each time slot is 1/2 of the cost for the time slot. The results are shown in Figure \ref{fig6}. \par
As can be seen from the figure, the optimal base bid oscillates significantly throughout the day. At certain times of the day, such as time slots 27, 43, 76, etc., in which users frequently interact with the ads, and the ad impressions at these time slots usually lead to click behavior. Hence, their optimal base bids are very high, close to the maximum bid of 300 set by iPinYou. During the midnight hours of the day, such as time slots 0-26, when users are not active, the optimal base bid is also relatively low. It is worth noting that the optimal base bid for some time slots is 0. This is because there is no click behavior for ad displays in these time slots. In theory, the agent should lower the base bid as much as possible to avoid wasting the budget on these low-value ad impressions. \par
There are significant differences in the base bids of the two RL-based bidding strategies. DRLB is constrained by the discrete action space of DQN and can only choose the adjustment value of the bidding factor for each time slot from a manually preset limited action space. The figure shows that such an adjustment has low sensitivity and is limited, leaving a large gap between the base bid and the optimal base bid of DRLB for most time slots. The FAB's base bid trend is basically the same as the optimal one. Since FAB relies on TD3's continuous action space, it can directly generate bidding factors for each time slot, allowing a more comprehensive range of adjustment for base bids and better fitting to optimal base bids. Therefore, the finer the granularity of the action space is, to a certain extent, the better. \par
\subsubsection{Impact of Reward Function Design}
\label{Impact on Reward Function Design}
The reward function is the driving force behind RL, which motivates the agent to make better action choices by evaluating the agent's behavior and guiding the value function updates. When using RL to optimize bidding strategies, the agent's goal is to find the bidding strategy that maximizes the advertiser's revenue. Therefore, an intuitive design of the reward function is to use the clicks generated by the winning ad impression as the immediate reward. However, in RTB, click behavior is very sparse. Using clicks as immediate rewards directly may result in the value function not being updated for a long time due to the absence of clicks, making the model learning speed significantly lower and weakening the convergence. In addition to clicks, pCTR is an essential measure of the ad valuation and can also serve as the immediate reward. In DRLB, the authors fit immediate rewards using a neural network whose inputs are state-action pairs and outputs are the corresponding rewards. In FAB, the authors propose a new way of reward shaping. Since the bids of FAB are adjusted based on LIN, the immediate reward is achieved by directly using the comparison results of the bid performance of FAB and LIN in the same time slot. \par
\begin{figure}[!]
	\centering
		\includegraphics[]{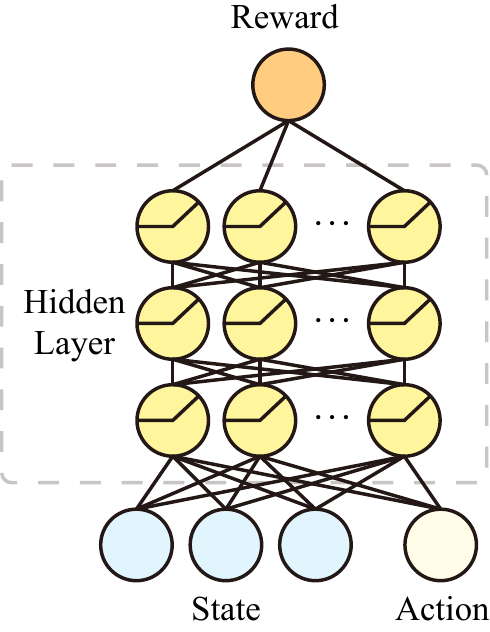}
	  \caption{Neural network structure of $Reward_{DNN}$.}\label{fig7}
\end{figure}

The above four reward functions are either derived from environmental feedback or tailored to their own models, and their design methods are summarized as follows:
\begin{itemize}
\item $Reward_{CLK}$: use the click behavior of winning ad impressions for the entire time slot as an immediate reward, as in formula \ref{equ15}
\begin{equation}
Reward_{CLK}=clks(t)
\label{equ15}
\end{equation}
\item $Reward_{PCTR}$: use the pCTR of winning ad impressions for the entire time slot as an immediate reward, as in formula \ref{equ16}
\begin{equation}
Reward_{PCTR}=pCTR(t)
\label{equ16}
\end{equation}
\item $Reward_{DNN}$: Using a fully connected neural network with three hidden layers and 100 neurons per layer to fit the immediate reward as in formula \ref{equ17}, the neural network structure is shown in Figure \ref{fig7}.
\begin{equation}
Reward_{DNN}=DNN(s_t,a_t)
\label{equ17}
\end{equation}
\item $Reward_{OP}$: Immediate rewards are given based on the number of clicks and cost compared to LIN over the time slot, as shown in formula \ref{equ18}. That is, if the results are better than those in LIN, the environment gives a positive reward to the bidding agent, else a negative reward.
\end{itemize}
\begin{equation}
\footnotesize
Reward_{OP}=
\begin{cases}
0.005, &if\ clks(t) \ge LIN\_clks(t)\ and\ cost(t) < LIN\_cost(t) \\
0.001, &if\ clks(t) \ge LIN\_clks(t)\ and\ cost(t) \ge LIN\_cost(t) \\
-0.0025, &if\ clks(t) < LIN\_clks(t)\ and\ cost(t) < LIN\_cost(t) \\
-0.005, &if\ clks(t) < LIN\_clks(t)\ and\ cost(t) \ge LIN\_cost(t) 
\end{cases}
\label{equ18}
\end{equation}
\par
The above reward functions either come from environmental feedback or are tailored to their models. To verify the generality of the four reward functions, we conduct experiments using FAB on dataset 1458 to compare the bidding performance of the four reward functions, and the results are shown in Table \ref{tab13}. $Reward_{OP}$ performs best in both budget conditions. It gets the most clicks and pCTR because this is the most suitable reward shaping method for FAB scenarios. Meanwhile, the performance of $Reward_{CLK}$ and $Reward_{PCTR}$ is notable. First of all, $Reward_{CLK}$ obtains a higher number of clicks under a sufficient budget, and $Reward_{PCTR}$ obtained a higher pCTR, which is deserved because these are the goals pursued by the two reward functions. In general, the pCTR of the ad impressions with click behaviors is also relatively high. Hence, the two reward functions' clicks and pCTR are close. However, when the budget is tight, $Reward_{PCTR}$ does not yield the same results as $Reward_{CLK}$ in both clicks and pCTR. $Reward_{PCTR}$ leads the bidding agent to spend many budgets on ad impressions to get the most immediate rewards. It will result in spending out the budget early when the budget is insufficient, causing bidding agents to lose all ad impressions with clicks in subsequent time slots and get a bad result. Therefore, the number of clicks is significantly lower, and the pCTR is less than that of $Reward_{CLK}$ due to the loss of ad display opportunities with clicks, which generally have a high pCTR. \par
As for $Reward_{DNN}$, we try to adjust the neural network parameters according to the FAB scenario. However, its bidding performance is not optimistic. In DRLB, $Reward_{DNN}$ fits the reward for the whole episode, while in FAB fits the reward at each time step and the new learning goal is more complex, leading to bad performance. Therefore, implementing the reward function using neural networks may require the reader to design and tune the parameters more carefully according to the usage scenario. \par
\begin{table}[!]
\small
\centering
\caption{Bidding results of different reward functions}
\begin{tabular}{ccccc}
\toprule
\toprule
\multirow{2}[0]{*}{\makecell*[c]{\textbf{Reward}\\ \textbf{Function}}} & \multicolumn{2}{c}{\textbf{1/2}} & \multicolumn{2}{c}{\textbf{1/16}} \\
\cmidrule{2-5}
& \textbf{clikcs} & \textbf{pCTR} & \textbf{clikcs} & \textbf{pCTR}\\
\midrule
$Reward_{CLK}$ & 445 & 521.4533 & 271 & 300.6337 \\
$Reward_{PCTR}$ & 443 & 523.0458 & 258 & 284.3149 \\
$Reward_{DNN}$ & 420 & 497.1420 & 231 & 233.2507 \\
$Reward_{OP}$ & 452 & 532.5847 & 276 & 305.0419 \\
\bottomrule
\bottomrule
\end{tabular}
\label{tab13}
\end{table}
In summary, the design of the reward function needs to be tailored to the local context to achieve optimal performance of the RL-based bidding strategy. If researchers do not want to spend too much time on the reward function design, it is also a good choice to directly use the click behavior of environmental feedback as an immediate reward. \par
\subsection{Suggestions}
\label{Suggestions}
Finally, we make some suggestions on how to optimize the bidding strategy in RTB based on the above experimental studies. First, the empirical analysis shows that the RL-based bidding strategy is well suited for the highly dynamic and uncertain RTB environment. Thus RL-based bidding strategy will replace the current linear bidding as the predominantly adopted bidding strategy on DSPs. Second, considering the partially observable characteristics of the DSP for the RTB environment, it is far simpler and more feasible to use the model-free RL algorithm to optimize the bidding strategy than the model-based RL algorithm. Third, in model-free RL, it is difficult for agents to directly learn the optimal bidding price for each impression, so similar to DRLB and FAB, it is good to let agents learn the bidding adjustment factor. \par
Furthermore, when designing the state, the primary consideration should be the budget constraint and preferably the relationship between the budget and the remaining ad delivery period, which can better reflect how the budget is spent over time. In addition, it is worthwhile to consider auction results from previous time slots. The design of the action should first consider continuous as a way to achieve fine-grained adjustments. The reward function ought to be designed according to the actual bidding strategy. For example, if the adjustment of bids is based on some baseline method, then a comparative reward function similar to FAB is a good choice. Further, it is generally more pervasive to use clicks on ad impressions as an immediate reward than the pCTR. \par
\section{Conclusion}
\label{Conclusion}
In RTB, advertisers determine the bids for ad impressions based on the ad valuation according to their bidding strategies. Therefore, the bidding strategy is crucial for advertisers to enhance the effectiveness of advertising campaigns. Previous researchers optimize bidding strategies using linear or pairwise methods. However, such static bidding strategies do not work well due to the highly dynamic nature of the auction environment. Recent research treats the advertiser’s bidding process as an MDP and uses RL to optimize the bidding strategy. In this paper, we experimentally perform a quantitative analysis of several representative bidding strategies. The experimental results show that the model-free RL-based framework for optimizing advertisers' real-time bidding strategies is indeed a very promising solution. So, we summarize the general steps for optimizing bidding strategies using RL algorithms. \par
\emph{Step 1:} Design a suitable bidding function for determining the bidding price for each ad impression that meets the targeting rules. The ideal bidding function consists of two parts: one is the base function, and the other is the bidding factor. The base function takes a linear or non-linear form and gives an initial bidding price depending on the valuation of the ad impression. The bidding factor, which is an adjustment value given by the bidding agent based on the level of competition in the auction environment and the budget usage, adjusts the initial bidding price and then for bidding. Introducing the bidding factor achieves the purpose of bidding adjustment with environment changes and transforms the bidding decision problem into a bidding factor decision problem. \par
\emph{Step 2:} Modeling the bidding factor decision process as an MDP. The state space's design needs to reflect the current environment's critical information accurately. For the RTB high-frequency trading model, we believe it is more practical to divide the state by time slot granularity than by single impression granularity. Action space is synergistic with the bidding function, which generates the bidding factor for each time slot. In general, continuous action space has better flexibility and scalability in generating bidding factors than discrete action space. Immediate rewards reflect the performance of bidding agents in a time slot. Therefore, the ideal reward function should consider the current state and action and combine it with the bidding results. \par
\emph{Step 3:} Select the appropriate RL algorithm to learn the optimal bidding strategy. Value-based RL, policy-based RL, and Actor-Critic algorithms, such as TD3, SAC, etc., are all feasible. In addition, we suggest a stochastic policy algorithm for scenarios like RTB where there is no unique optimal action, which may yield higher learning efficiency. \par
\emph{Step 4:} Use the optimal bidding strategy to participate in the bidding in the new ad delivery period and update the bidding strategy iteratively based on the actual transaction records and user feedback. \par
In summary, the above steps are only a preliminary summary of our experience and methodology for optimizing bidding strategies using RL based on a large number of experiments completed. We believe that through the efforts of more experts in the field, we can form a framework for an effective bidding strategy for the highly dynamic RTB environment. Finally, we discuss the future research hotspots for bidding strategies, focusing on the following three areas. \par 
\textbf{State design.} The state definition of existing RL-based bidding strategies is based on time slots. The state of each time slot is represented by the bidding results of the previous time slot. Whether such a state representation is clear for the agent needs to be further verified. Moreover, we currently only use the bidding result of the previous time slot to represent the state, ignoring the action and do not consider the state information of earlier time slots and the corresponding actions and rewards. It is usually helpful to include historical information as a basis for decision-making. In addition to the temporal stacking of state information across multiple time slots, recurrent neural networks can be used to exploit historical information further. \par
\textbf{Reward function design.} Another challenge for MDP modeling is the definition of the reward function. In RTB, it is uncertain whether the bidding agent will earn actual revenue (clicks or conversions) after bidding. Existing RL-based bidding strategies use the pCTR of winning impressions as the immediate reward and do not yield the desired results in the empirical analysis. FAB uses a comparative approach with the baseline to define reward function. It is a novel approach and achieves good results but does not have generalization capabilities. In future research, a general form of definition for the reward function in RTB would speed up the implementation of RL-based bidding strategies in the industry. It is an excellent choice to learn the representation of the reward function with the help of inverse reinforcement learning. \par
\textbf{Study of RL algorithms conforming to RTB scenarios.} Current RL-based bidding strategies use proven algorithms such as DQN, TD3, etc. These algorithms are successful in gaming and robot control but are unstable in convergence in RTB scenarios. Therefore, RL algorithms applicable to RTB scenarios will also become a research hotspot.
\section*{Acknowledgements}
This work was supported by the National Natural Science Foundation of China under Grant 61202445; the Fundamental Research Funds for the Central Universities under Grant ZYGX2016J096.
\bibliographystyle{elsarticle-num}

\bibliography{RLBidEA}

%
%
\end{document}